\documentclass[letter]{aa} 

\usepackage{graphicx}
\usepackage{txfonts}
\usepackage{natbib}
\bibpunct{(}{)}{;}{a}{}{,}

\usepackage{lscape}
\usepackage{longtable}
\usepackage{amssymb}

\def\kms{km~s$^{-1}$}
\def\g2301{G023.01$-$00.41}

\begin{document}

%
%
   \title{Velocity and magnetic fields within 1000 AU from a massive YSO}

   \author{A. Sanna\inst{1} \and G. Surcis\inst{2} \and L. Moscadelli\inst{3} \and R. Cesaroni \inst{3}
            \and C. Goddi\inst{4} \and W.H.T. Vlemmings \inst{5}  \and A. Caratti~o~Garatti \inst{6}}
            

   \institute{Max-Planck-Institut f\"{u}r Radioastronomie, Auf dem H\"{u}gel 69, 53121 Bonn, Germany\\
   \email{asanna@mpifr-bonn.mpg.de}
   \and JIVE, Joint Institute for VLBI in Europe, Postbus 2, 7990 AA Dwingeloo, The Netherlands
   \and INAF, Osservatorio Astrofisico di Arcetri, Largo E. Fermi 5, 50125 Firenze, Italy
   \and Department of Astrophysics/IMAPP, Radboud University Nijmegen, PO Box 9010, NL-6500 GL Nijmegen, the Netherlands
   \and Department of Earth and Space Sciences, Chalmers University of Technology, Onsala Space Observatory, SE-439 92 Onsala, Sweden
   \and Dublin Institute for Advanced Studies, School of Cosmic Physics, Astronomy \& Astrophysics Section, 31 Fitzwilliam Place, Dublin 2, Ireland}

   \date{Received June, 2015; accepted September, 2015}


  \abstract
   {}
   {We want to study the velocity and magnetic field morphology in the vicinity (<1000\,AU) of a massive young stellar object (YSO), at very high 
   spatial resolution (10--100\,AU).}
   {We performed milli-arcsecond polarimetric observations of the strong CH$_3$OH maser emission observed in the vicinity of an O-type YSO, in G023.01$-$00.41.
    We have combined this information with the velocity field of the CH$_3$OH masing gas previously measured at the same angular resolution. We analyse the 
    velocity and magnetic fields in the reference system defined by the direction of the molecular outflow and the equatorial plane of the hot molecular core
    at its base, as recently observed on sub-arcsecond scales.}
   {We provide a first detailed picture of the gas dynamics and magnetic field configuration within a radius of 2000\,AU from a massive YSO. We have
   been able to reproduce the magnetic field lines for the outer regions (>600\,AU) of the molecular envelope, where the magnetic field orientation shows a
   smooth change with the maser cloudlets position ($0.2\degr$\,AU$^{-1}$). Overall, the velocity field vectors well accommodate with the local, magnetic field
   direction, but still show an average misalignment of $30\degr$. We interpret this finding as the contribution of a turbulent velocity field of about 
   3.5\,\kms, responsible for braking up the alignment between the velocity and magnetic field vectors. We do resolve different gas flows which develop
   both along the outflow axis and across the disk plane, with an average speed of 7\,\kms. In the direction of the outflow axis, we establish a collimation of
   the gas flow, at a distance of about 1000\,AU from the disk plane. In the disk region, gas appears to stream outward along the
   disk plane for radii greater than 500--600\,AU, and inward for shorter radii.}
   {}

   \keywords{ISM: kinematics and dynamics --
             Masers --
             Stars: formation --
             Stars: individual: G023.01$-$00.41
             }

   \maketitle
%

\section{Introduction}

The role of magnetic fields in regulating the gas dynamics in the vicinity of growing, massive, 
young stellar objects (YSOs) is still a matter of debate (e.g., \citealt{Crutcher2010,Zhang2014}). Recent magneto-hydrodynamics (MHD)
models, simulating the build-up of massive protostars in the inner few 1000\,AU, have shown that magnetic fields
may contribute significantly (1) to the degree of outflow collimation, and (2) to stabilizing both Keplerian
and sub-Keplerian disks against fragmentation (e.g., \citealt{Seifried2011,Seifried2012}).  
In this context, Very Long Baseline Interferometry (VLBI) observations of maser emission, arising within a few
1000\,AU from massive YSOs, allow us to determine both the velocity distribution and the magnetic field
configuration close to the accreting protostar (e.g., \citealt{Sanna2010a,Sanna2010b,Goddi2011,Moscadelli2011,Surcis2015}).
This gives us the unique chance to investigate, at a 10--100\,AU scale, whether or not the magnetic field influences the gas kinematics. 

\g2301\,  is a luminous star-forming region of about $\rm 4\times10^4\,L_{\odot}$ \citep{Sanna2014},
located at a trigonometric distance of 4.6\,kpc \citep{Brunthaler2009}. This
star forming site harbors a flattened, hot molecular core (HMC) which is centered on an active site of 
strong maser and radio continuum emission \citep[their Fig.\,4]{Sanna2010b}. The kinematics  
of warm (200\,K) gas in the inner 3000\,AU, as traced with CH$_3$CN and thermal CH$_3$OH lines, shows 
the composition of two, orthogonal, velocity fields \citep[their Fig.\,3]{Sanna2014}. The velocity component
which dominates at larger scales is aligned with the axis of a collimated bipolar outflow, traced progressively away 
from the HMC center with SiO and CO gas emission. Since the outflow emission is almost perpendicular to the line
of sight, any associated disk should be seen edge-on, which makes this object an excellent target to study
the gas dynamics in the vicinity of an O-type YSO. Furthermore, the 3D gas kinematics revealed by the CH$_3$OH
masers shows a funnel-like morphology \citep[their Fig.\,6]{Sanna2010b}, which was best interpreted as the base
of the outflow cavity (or the surface of a flared disk) with a size between 1000~and 2000\,AU. 

With this in mind, we decided to use the synergy between maser proper motions and polarization measurements,
targeting the rich CH$_3$OH maser spectrum observed in \g2301, to investigate whether magnetic fields may be
actively driving the circumstellar gas motion around a massive YSO. That can be assessed by quantifying whether a correlation
exists between the orientation of the velocity and polarization vectors locally, as measured for individual CH$_3$OH
masing cloudlets on scales of a few AU. In order to compare the magnetic field orientation with the velocity field previously
measured by \citet{Sanna2010b},  we conducted polarimetric observations of the 6.7\,GHz CH$_3$OH masers towards
\g2301 with the European VLBI Network (EVN\footnote{The European VLBI Network is a joint facility of European,
Chinese, South African and other radio astronomy institutes funded by their national research councils. }).

\onlfig{
\begin{figure*}
\centering
\includegraphics [angle= 0, scale=0.6]{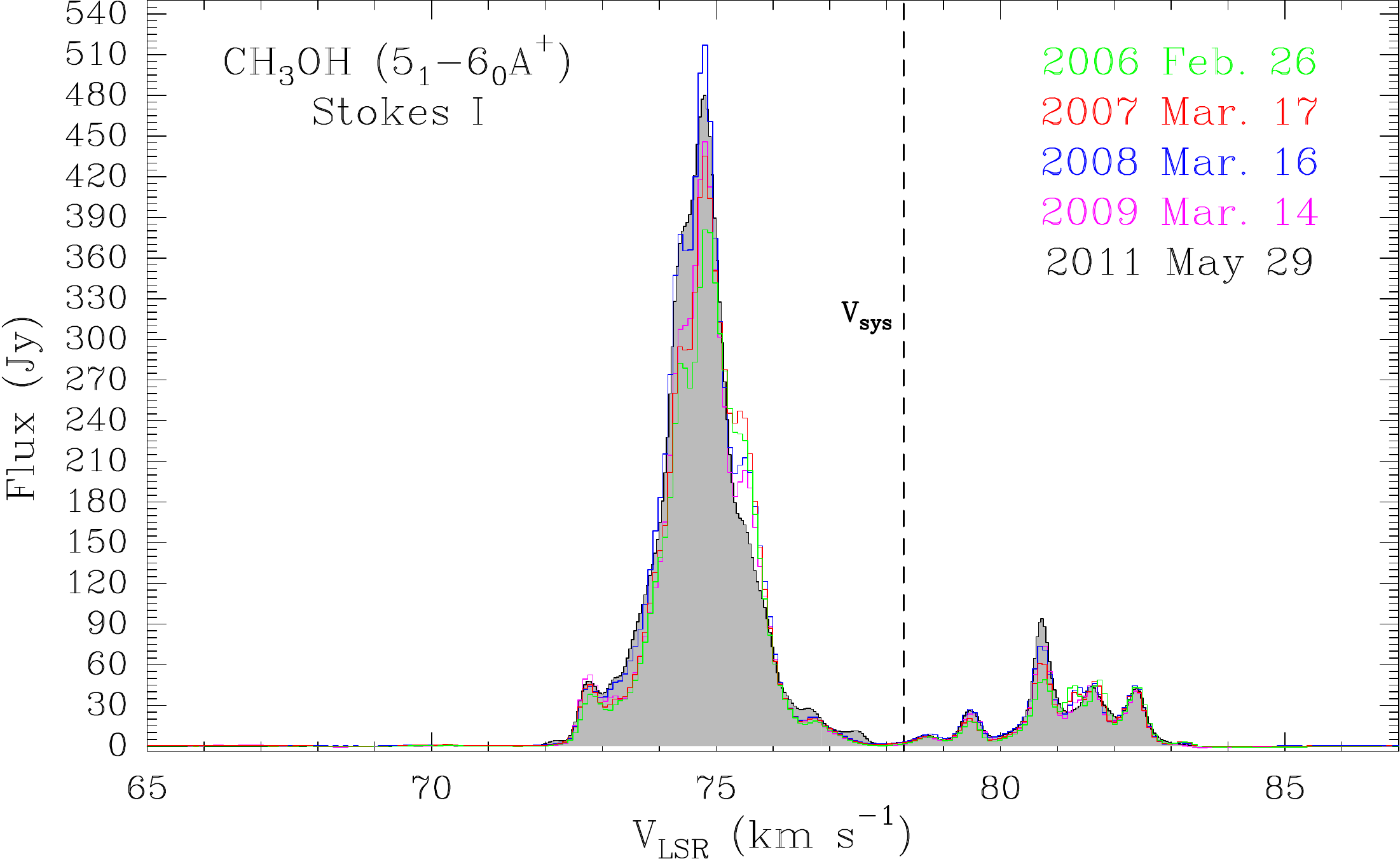}
\caption{Effelsberg total-power spectra toward \g2301, obtained from EVN observations at C-band, over 5\,yr (from \citealt{Sanna2010b},
and current measurements). Observing dates are indicated on the top right of the plot. The dotted vertical line marks the systemic velocity
(V$_{sys}$) of the HMC, as inferred from CH$_3$CN measurements.}
\label{fig1}
\end{figure*}
}

\section{Observations and Calibration}

We employed the EVN to observe in full polarization mode the $5_1-6_0$\,A$^+$ CH$_3$OH maser transition, at the
rest frequency of 6668.519\,MHz, toward \g2301. The observations were conducted under program ES067 on 2011 
May 29. We made use of a single frequency setup to obtain both a high spectral sampling (0.98\,kHz) of the maser lines,
and a bandwidth large enough (2\,MHz) to accurately measure the continuum emission of the calibrator, J\,2202$+$4216. 
This calibrator served both as a fringe finder and polarization calibrator, and was observed every 45\,min to properly calibrate
the polarization leakage. Since many maser features are expected to be linearly polarized at a level of about 1\%, to reach a
conservative detection above $5\,\sigma$ over half of the maser cloudlets previously detected (with peak intensities 
$>$3\,Jy\,beam$^{-1}$), we spent about 4.5 hours on-source. The EVN data were processed with the SFXC software
correlator \citep{Keimpema2015} at the Joint Institute for VLBI in Europe by using an averaging time of 2\,s. The single-dish
spectrum of the CH$_3$OH maser emission toward \g2301 is plotted in Fig.\,\ref{fig1}.

Data were reduced with the NRAO Astronomical Image Processing System (AIPS). We mapped the 
CH$_3$OH maser distribution with a (robust 0) beam size of $\rm 11\,mas\times4\,mas$, achieving a thermal noise 
of 5\,mJy\,beam$^{-1}$. To calibrate the systematic rotation of the linear polarization angle ($\chi_{\rm pol}$) in the
EVN dataset, we compared the EVN measurement of $\chi_{\rm pol}$ obtained on J\,2202$+$4216 with two, consecutive, VLA
polarimetric observations of the same calibrator\footnote{http://www.aoc.nrao.edu/~smyers/evlapolcal/polcal$\_$master.html}
bracketing our VLBI observations (on 2011 April 30, and 2012 February 3). The linear polarization angle of J\,2202$+$4216
estimated with the VLA remained nearly constant with an average value of $-31\degr \pm 1\degr$ (position angles, e.g., $\chi_{\rm pol}$,
are measured east of north, unless otherwise stated). Therefore, the $\chi_{\rm pol}$ measurements obtained with the
EVN dataset are affected by a systematic uncertainty of no more than a few degrees. The uncertainty of $\chi_{\rm pol}$ due to
thermal noise was obtained from the relative error of the polarization intensity measurement, following \citet{Wardle1974}. Details
about the polarization calibration can be found in \citet{Surcis2013}.

\begin{figure}
\centering
\includegraphics [angle= 0, width=\hsize]{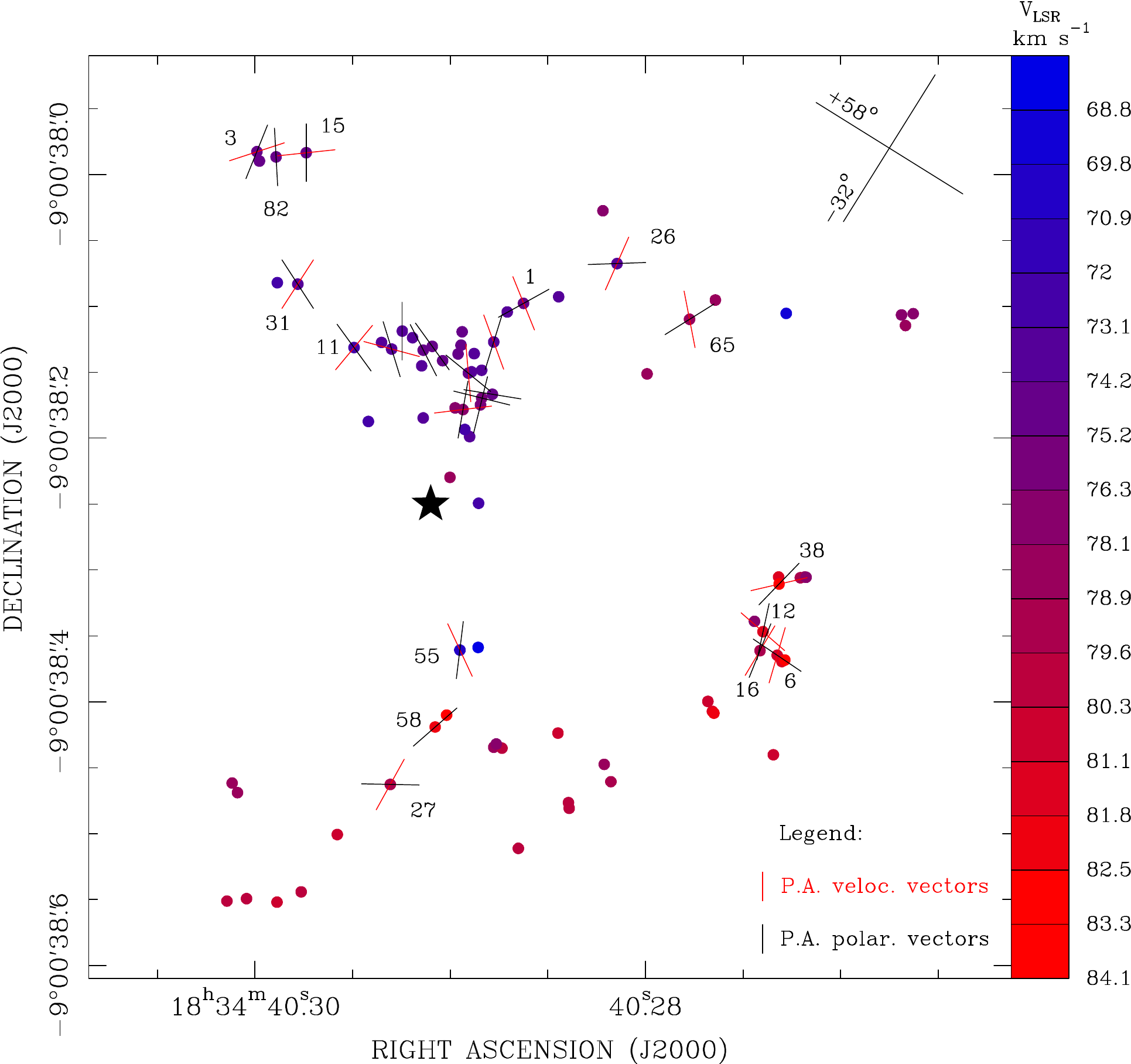}
\caption{Comparison of the velocity and polarization vectors across the CH$_3$OH maser distribution
in \g2301. We show the distribution of the 6.7\,GHz maser cloudlets (dots) within a field-of-view of
about 2000\,AU, following \citet[their Figure~6]{Sanna2010b}. Maser velocities along the line-of-sight
are color coded according to the right-hand V$_{\rm LSR}$ scale. Lines drawn on individual maser cloudlets
indicate the orientation of the velocity (red) and polarization (black) fields locally, as obtained from
VLBI measurements. \emph{For clarity, velocity vectors are only shown for cloudlets with an associated
polarization vector.} For isolated maser cloudlets, numbers correspond to the maser labels given in
Tab.~\ref{tab1}. The star marks the peak position of the high-excitation, CH$_3$OH, thermal line detected
by \citet{Sanna2014}, which is the origin of the reference system (e.g., upper right corner) used
in Fig.\,\ref{fig3} and~\ref{fig4}.}
\label{fig2}
\end{figure}

\section{Results}\label{results}

In the HMC center of \g2301, the milli-arcsecond distribution of the 6.7\,GHz CH$_3$OH maser cloudlets has not changed 
over the five years spanned by our observations. Although at some maser velocities the overall flux density has smoothly
changed in time  (Fig.\,\ref{fig1}), this variation affects the flux density of individual cloudlets, whereas the overall maser distribution
is preserved. Among the eighty maser cloudlets detected by \citet{Sanna2010b}, only about one third shows linearly 
polarized emission, and their properties are listed in Tab.\,\ref{tab1}. Given that we know only the orientation of the linear polarization
vectors ($\chi_{\rm pol}$), while their direction is undefined, we folded these values in the range $-90\degr$<$\chi_{\rm pol}$<$90\degr$.
The linear polarization fraction detected among the maser cloudlets ranges between 0.6\% and 9.2\%, with an average value of 2\%.
The orientation of the linear polarization vectors is superposed on the CH$_3$OH maser distribution in Fig.\,\ref{fig2}.  For cloudlets
with detected linear polarization, we also draw the direction  ($\chi_{\rm vel}$) of the velocity vectors for comparison.

\begin{figure}
\centering
\includegraphics [angle= 0, scale= 1.0]{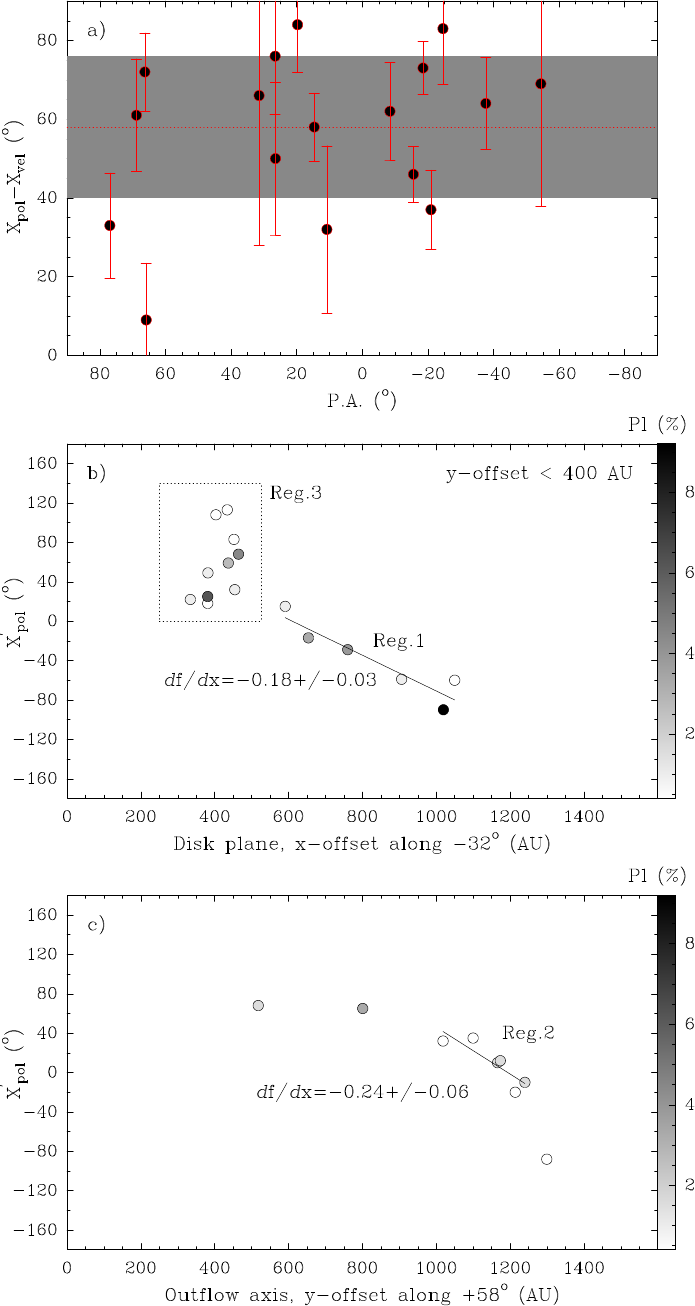}
\caption{Analysis of the polarization vectors orientation ($\chi_{\rm pol}$) as a function of the maser cloudlet position
(see Sect.\,\ref{results}).  
\textbf{a)} plot of the difference (and uncertainty) between the orientations of the velocity ($\chi_{\rm vel}$)
and polarization ($\chi_{\rm pol}$) vectors as a function of the sky position angle (P.A.) of each cloudlet (defined in
Sect.\,\ref{results}). Cloudlets with P.A. in the range $90\degr$--$270\degr$ have been folded
in the range $-90\degr$<P.A.<$90\degr$. The grey area marks the boundary of the weighted standard deviation
($\pm 1\,\sigma$) for the $\chi_{\rm pol}$--$\chi_{\rm vel}$ values.
\textbf{b)} plot of $\chi'_{\rm pol}$ as a function of the projected distance along the equatorial plane of the  
HMC ($-32\degr$). $\chi'_{\rm pol}$ gives the values of $\chi_{\rm pol}$ as measured counterclockwise starting 
from a sky position angle of $-32\degr$. In this panel, only points with offsets less than 400\,AU from the equatorial plane
have been plotted.  For each $\chi'_{\rm pol}$ value, the polarization fraction is quantified according to the wedge on the
right-hand side. The uncertainty of  the $\chi'_{\rm pol}$ measurements is of the same order as the marker size. The
different regions identified in Sect.\,\ref{results} are also indicated. 
\textbf{c)} similar to panel \textbf{b)} with projected distances along the outflow axis  ($58\degr$). 
Data points plotted in panel \textbf{b)} are not drawn. }
\label{fig3}
\end{figure}

In Fig.\,\ref{fig3}, we study the distribution of the polarization vectors orientation with respect to the position of the 
CH$_3$OH maser cloudlets. In Fig.\,\ref{fig3}a, we plot the minimum difference ($\chi_{\rm pol}$--$\chi_{\rm vel}$)
between the orientation of the polarization and velocity vectors as a function of the sky position angle of each cloudlet. This
position angle is measured east of north with respect to the HMC center (star symbol in Fig.\,\ref{fig2}), which is defined as
the peak position of the high-excitation, CH$_3$OH, thermal emission at the systemic velocity of the core \citep{Sanna2014}. 
This position represents the current best estimate of the YSO position.
The $\chi_{\rm pol}$--$\chi_{\rm vel}$ distribution has a weighted average of $58\degr$, and a weighted dispersion of $\pm 18\degr$
(grey area in Fig\,\ref{fig3}a). In Fig.\,\ref{fig3}b and\,c, maser positions are projected along the two orthogonal axes defined by the
direction of the molecular outflow ($+58\degr$), and that of the elongated HMC ($-32\degr$), with the origin at the HMC center.
Assuming a perfect symmetry of the gas dynamics with respect to these two axes, we produce a mirror image 
of the linear polarization vectors of each cloudlet on a single quadrant ($\chi'_{\rm pol}$). For the $\chi'_{\rm pol}$ values,
we also try to solve the ambiguity of $\pm180\degr$, assuming that cloudlets close in space would show a smooth change of
the polarization (and magnetic) field with position. A posteriori, this criterion is found to minimize the difference between
the direction of the linear polarization vectors and that of the corresponding velocity vectors. In Fig.\,\ref{fig3}b and\,c, we identify
two regions, labeled Reg.\,1 and\,2,  where $\chi'_{\rm pol}$ changes smoothly with the projected distance of the maser cloudlets.
These regions are fitted by a linear slope of about $0.2\degr$\,AU$^{-1}$, and correspond to nearby cloudlets which also show a
smooth variation of the velocity field (Fig\,\ref{fig4}). A third region labeled Reg.\,3, shows a variation of $\chi'_{\rm pol}$ by more
than $100\degr$ over a small range of projected distances. This region corresponds to cloudlets with the shortest projected distances,
and will be discussed further in Sect.\,\ref{discussion}.

\section{Discussion}\label{discussion}

\begin{figure*}
\sidecaption
\includegraphics [angle= 0, scale= 0.5]{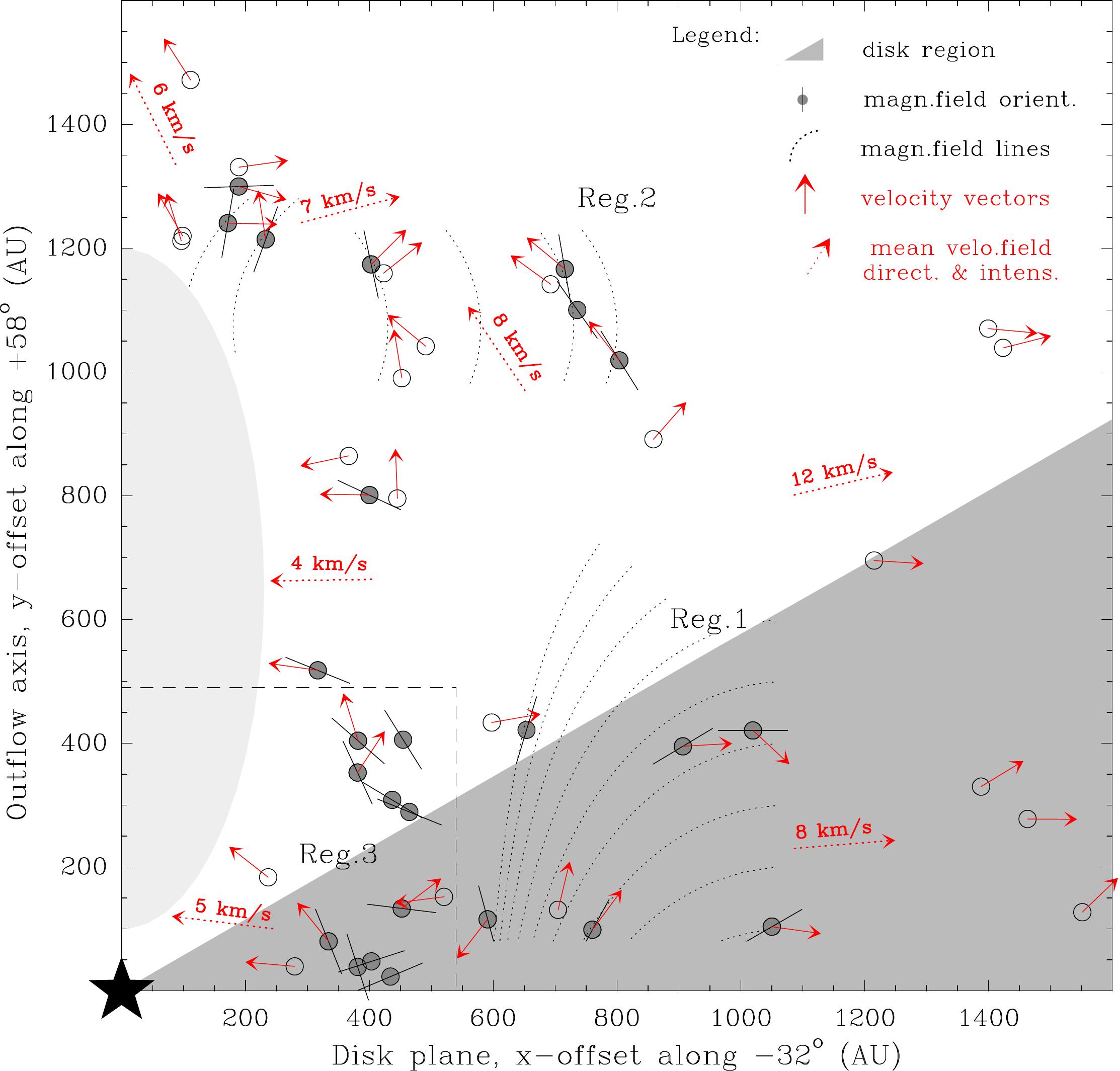}
\caption{Combined plot of the velocity and magnetic field measurements obtained from the 6.7\,GHz, CH$_3$OH, maser cloudlets
(dots) in  \g2301. The axes are aligned with the outflow and disk directions, as defined by the sub-arcsecond SMA
observations in the 1\,mm band from \citet{Sanna2014}. The origin is set at the HMC center, as defined in Sect.\,\ref{results} (star
symbol in Fig.\,\ref{fig2}). The dark grey area marks the region where emission from a flared disk with axes ratio of $1/2$ is expected. 
The light grey area marks the region close to the outflow axis which is devoid of maser emission.  
For each maser cloudlet, the local directions of the velocity vector (red solid arrows) and/or that of the magnetic field vector (black solid lines)
are reported where available. Empty dots are used for cloudlets which do not show polarized emission. The local average of both the magnitude
and direction of the velocity field is indicated by red dotted arrows. These values were derived by grouping close by features with similar
velocity directions.
The different regions identified through the polarization vectors analysis (Fig.\,\ref{fig3}), and labeled from 1 to 3, are indicated.
For regions showing a smooth variation of the polarization (and magnetic) field vectors with the maser cloudlets position, we plot the families
of curves (black dotted lines) which best match the tangent to the local, magnetic field measurements (see Sect.\,\ref{discussion}). The
pattern of these families of curves approximates the spatial morphology of the magnetic field lines locally. The direction of the velocity vectors
is on average tilted by $30\degr$ with respect to the local, magnetic field direction.}
\label{fig4}
\end{figure*}

To infer the local magnetic field orientation, we ran the radiative transfer model by \citet{Vlemmings2010} for each cloudlet 
with detected linearly polarized emission (last columns of Tab.\,\ref{tab1}). According to the output 
parameters of this modeling ($\theta$\,$\gg$\,$55\degr$), the magnetic field orientation is perpendicular to the
polarization vectors for all maser components. This information is used in  Fig.\,\ref{fig4} to plot
the local, magnetic field orientation by rotating the $\chi'_{\rm pol}$ values by $90\degr$.  
More details about the radiative transfer modeling used, can be found in \citet{Surcis2013} .

In Fig.\,\ref{fig4}, we make use of the outflow/disk geometry described in \citet{Sanna2014}, to give 
a complete picture of the velocity and magnetic fields within 1000 AU from the HMC center. In this plot,
we produce a mirror image of all the measurements of velocity and magnetic field vectors obtained from the 
6.7\,GHz, CH$_3$OH, maser cloudlets, as if they were sampling a single quadrant defined by the outflow
direction and the disk plane. This picture holds under the assumption that the gas dynamics shows a symmetric
behavior with respect to the outflow axis and the disk plane.  Given the uncertainty of $30\degr$ on the disk
inclination  \citep{Sanna2014}, one should keep in mind that the maser cloudlets might be closer to the disk plane
than they appear. The high-density molecular tracers observed toward \g2301\, show a fairly constant ratio
of $\sim$2 between the major and minor axis of the HMC. In Fig.\,\ref{fig4}, this ratio is interpreted as if it was
due to a flared disk with a semi opening angle of $30\degr$ (dark grey area). We also mark a central region along
the outflow axis which is devoid of  CH$_3$OH maser emission (light grey area).

In an attempt to derive a continuous, magnetic field morphology, which reproduces the local maser measurements,
we considered those regions (Reg.\,1 and~2) showing a smooth change of the polarization vectors (and magnetic field)
orientation with the maser cloudlets position. At a first order, these slopes have been approximated by a linear fit as
shown in Fig.\,\ref{fig3} ($\chi'_{\rm pol}$=$f(x)$). We can then integrate the tangent of $[f(x)+90\degr]$, in order to
derive the families of curves which best fit the local, magnetic field orientation at the maser cloudlets position. These
curves give a first order representation of the local morphology of the magnetic field lines (black dotted lines in Fig\,\ref{fig4}).

The velocity field traced by maser cloudlets belonging to Reg.\,1 and~2, provides a consistent picture of gas outflowing from 
the HMC center along the magnetic field lines. In Reg.\,1, for small heights over the disk plane (<400\,AU), the velocity field well
accommodates with the magnetic field lines, starting from projected distances of 600\,AU up to about 1100\,AU. 
Further away, the velocity field mainly expands parallel to the disk plane, and shows the highest maser velocities ($\sim$10\,\kms).
On the other hand, maser cloudlets belonging to Reg.\,2 and upwards, expand and get collimated in the direction of the
outflow axis. In particular, as one proceeds upwards along the magnetic field lines, and closer to the outflow axis, both the velocity
and magnetic field vectors independently undergo a turn of $90\degr$. This feature may be interpreted as a result of the
complex gas dynamics where both a slow and fast velocity component exist (e.g., \citealt{Seifried2012}, their Fig.\,12). Indeed,
at about 2000\,AU along the outflow axis (not shown in Fig.\,\ref{fig4}), a shock front of dense gas traced by H$_2$O masers 
shows gas velocities of 20\,\kms \citep[their Fig.\,5b]{Sanna2010b}, three times higher than those traced by the CH$_3$OH  
gas in Reg.\,2.
   
Methanol maser cloudlets belonging to Reg.\,3 have projected distances of less than about 500\,AU (dashed box in Fig.\,\ref{fig4}),
both along the disk plane and the outflow axis. In this inner region, the velocity field is composed of (at least) two different
motions, 1) an inflowing motion closer to the disk plane (y-offset\,$<$\,200\,AU), and 2) an upward motion for higher offsets.
The magnetic field pattern of this region is more complex, and we did not attempt to reproduce the magnetic field lines.
Still, nearby cloudlets show a similar orientation of the magnetic field vectors, and confirm the accuracy of our measurements.

Interestingly, Reg.\,3 corresponds to a diffuse halo emission from strong CH$_3$OH masers \citep[see their Fig.\,6b]{Sanna2010b}, which are
likely saturated due to the vicinity of the central IR source. This ridge of extended emission is significantly elongated in the direction of the
outflow axis, in agreement with the average orientation of the magnetic field vectors between y-offset$\sim$300 and 400\,AU, and the upward motions
detected there. This evidence makes us speculate that Reg.\,3 may trace the outer launching region of the primary outflow, in agreement with 
recent MHD simulations by \citet[their Fig.\,5]{Seifried2012}. Furthermore, we make use of the inward stream of gas close to the
disk plane, to obtain an estimate of the mass inflow rate, 
$\rm \dot{M}_{in} = (5.0\times10^{-5}\,M_{\odot}\,yr^{-1})\,R_{100}^{2}\,v_{10}\,n_{8}$. 
In this formula, $\rm R_{100}$, $\rm v_{10}$, and $\rm n_{8}$ are the mean radius of the inward stream in units of 100\,AU,
its velocity in units of 10\,km\,s$^{-1}$, and the volume density of molecular hydrogen in units of $\rm 10^8\,cm^{-3}$, respectively.
We take into account that the inward stream of gas is confined within an angle of $60\degr$ either side of the YSO, and allow for gas densities 
as high as $\rm 10^8\,cm^{-3}$, above which the Class\,{\sc ii} CH$_3$OH masers start to be quenched \citep{Cragg2005}. Noticeably, we find no
maser detection at closer distances to the HMC center. At an average distance of 300\,AU, the inward stream of gas, flowing at a velocity of
$\rm 5\,km\,s^{-1}$, brings a mass inflow rate of $\rm 2\times10^{-4}\,M_{\odot}\,yr^{-1}$. 

We finally consider the average misalignment of $60\degr$ observed between the velocity and polarization vectors (Fig.\,\ref{fig3}a),
which translates to $30\degr$  between the velocity and magnetic field vectors (Fig.\,\ref{fig4}). If gas and magnetic field were fully coupled,
we would observe gas flowing along the magnetic field lines, and expect an average misalignment close to zero. Given that the observed
misalignment appears randomly distributed ($1\,\sigma$\,=\,$\pm 18\degr$) about  $30\degr$, we model this effect as it was due to a
random (turbulent) velocity component, which adds to ordered velocity vectors aligned with the magnetic field lines. By considering a
velocity vector of 7\,\kms, as averaged across the whole region, we estimate this turbulent contribution to be of the order of 3.5\,\kms. 
This value is very similar to the velocity dispersion (4--5\,\kms) derived from the CH$_3$CN linewidth in the inner 3000\,AU from the HMC center
\citep{Sanna2014}, which fairly supports our estimate.

\begin{acknowledgements}

Financial support by the Deutsche Forschungsgemeinschaft (DFG) Priority Program 1573 is gratefully acknowledged.
WV acknowledges financial support from the European Research Council through ERC consolidator grant 614264.
A.C.G. was supported by the Science Foundation of Ireland, grant 13/ERC/I2907.
\end{acknowledgements}


\bibliographystyle{aa}
\bibliography{asanna2206}


\onltab{
\begin{table*}
\caption{\label{tab1} Parameters of 6.7\,GHz methanol maser cloudlets with detected linear polarization.}

\centering
\begin{tabular}{r r r c r r c c c}
\hline
\hline
 & & & & & & \multicolumn{3}{c}{Radiative Transfer Model} \\
\multicolumn{1}{c}{Feature} & \multicolumn{1}{c}{V$_{\rm LSR}$} & \multicolumn{1}{c}{F$_{\rm peak}$} & P$_{\ell}$ &
\multicolumn{1}{c}{$\chi_{\rm pol}$} & \multicolumn{1}{c}{$\chi_{\rm vel}$} & $\rm \Delta V_{\rm i}$ & T$_{\rm b}\Delta\Omega$ & $\theta$  \\
\multicolumn{1}{c}{\#} & \multicolumn{1}{c}{(km\,s$^{-1}$)} & \multicolumn{1}{c}{(Jy\,beam$^{-1}$)} & (\%) & \multicolumn{1}{c}{($\degr$)} &
\multicolumn{1}{c}{($\degr$)} & (km\,s$^{-1}$) & (log\,K\,sr) & ($\degr$) \\
\hline
 & & & & & & & & \\

\multicolumn{9}{c}{\textbf{Northern Region -- Dec.\,(J2000)\,$>-$9:00:38.30}} \\

1    & 74.76 &105.50 & $ 4.0 \pm 0.1$ & $ -61 \pm 2  $ & $ +22 \pm 14$ & $1.5^{+0.2}_{-0.5}$   & $ 9.3^{+0.4}_{-0.1}$ & $88^{+2}_{-47}$  \\
2    & 75.55 & 34.52  & $ 1.1 \pm 0.1$ & $ -14 \pm 1  $ & ...           & $2.6^{+0.1}_{-0.3}$   & $ 8.6^{+0.2}_{-0.2}$ & $89^{+1}_{-23}$  \\
3    & 74.32 & 44.47  & $ 2.2 \pm 0.3$ & $ -22 \pm 4  $ & $+108 \pm 19$ & $1.4^{+0.2}_{-0.4}$   & $ 9.0^{+0.5}_{-0.1}$ & $79^{+10}_{-41}$ \\
4    & 74.54 & 97.84  & $ 1.1 \pm 0.1$ & $ +51 \pm 1  $ & $  +5 \pm  7$ & $2.2^{+0.2}_{-0.3}$   & $ 8.6^{+0.2}_{-0.1}$ & $90^{+25}_{-25}$ \\
5    & 74.81 & 45.39  & $ 0.9 \pm 0.2$ & $ +81 \pm 2  $ & ...           & $2.1^{+0.2}_{-0.3}$   & $ 8.6^{+0.5}_{-0.2}$ & $78^{+12}_{-39}$ \\
7    & 75.86 & 38.56  & $ 1.1 \pm 0.1$ & $ +76 \pm 1  $ & ...           & $2.3^{+0.2}_{-0.2}$   & $ 8.6^{+0.3}_{-0.1}$ & $84^{+6}_{-40}$  \\
11  & 72.70 & 30.90  & $ 2.2 \pm 0.1$ & $ +36 \pm 5  $ & $+140 \pm 14$ & $2.1^{+0.4}_{-0.4}$   & $ 9.0^{+0.3}_{-0.1}$ & $90^{+44}_{-44}$ \\
13  & 74.41 &  8.24   & $ 4.6 \pm 0.8$ & $ +36 \pm 2  $ & ...           & $1.1^{+0.2}_{-0.4}$   & $ 9.4^{+0.5}_{-0.3}$ & $79^{+11}_{-16}$ \\
14  & 74.32 & 19.42  & $ 3.3 \pm 0.3$ & $ +27 \pm 1  $ & ...           & $1.4^{+0.2}_{-0.5}$   & $ 9.2^{+0.4}_{-0.1}$ & $82^{+7}_{-41}$  \\
15  & 74.67 & 11.37  & $ 0.9 \pm 0.3$ & $     0 \pm 9  $ & $ +96 \pm  8$ & $1.4^{+0.2}_{-0.2}$   & $ 8.5^{+0.4}_{-0.6}$ & $74^{+15}_{-38}$ \\
18  & 73.75 &  9.48   & $ 1.7 \pm 0.3$ & $ +17 \pm 5  $ & $ +75 \pm  7$ & $1.4^{+0.2}_{-0.3}$   & $ 8.9^{+0.4}_{-0.3}$ & $80^{+10}_{-39}$ \\
19  & 73.49 & 10.69  & $ 1.4 \pm 0.1$ & $  -17 \pm 1  $ & $-160 \pm 10$ & $2.2^{+0.2}_{-0.3}$   & $ 8.7^{+0.3}_{-0.1}$ & $84^{+6}_{-41}$  \\
21  & 76.60 & 10.09  & $ 1.4 \pm 0.3$ & $  -10 \pm 3  $ & $ +97 \pm  6$ & $2.4^{+0.3}_{-0.3}$   & $ 8.7^{+0.2}_{-0.9}$ & $90^{+23}_{-23}$ \\
24  & 73.93 &  4.03   & $ 1.5 \pm 0.3$ & $     0 \pm 12 $ & ...           & $<0.5$                & $ 8.9^{+0.2}_{-0.4}$ & $84^{+6}_{-19}$  \\
26  & 73.18 & 11.29  & $ 0.6 \pm 0.2$ & $  -88 \pm 6  $ & $ -24 \pm 10$ & $1.3^{+0.1}_{-0.2}$   & $ 8.4^{+0.4}_{-0.4}$ & $75^{+13}_{-40}$ \\
31  & 73.27 &  4.32   & $ 3.8 \pm 0.4$ & $ +33 \pm 2  $ & $+147 \pm 38$ & $1.6^{+0.3}_{-0.5}$   & $ 9.3^{+0.3}_{-0.1}$ & $84^{+6}_{-40}$  \\
65  & 78.71 &  0.82   & $ 9.2 \pm 0.4$ & $  -58 \pm 2  $ & $ +11 \pm 31$ & $<0.5$                & $10.0^{+0.1}_{-0.2}$ & $90^{+7}_{-7}$   \\
82\tablefootmark{a}   & 74.85 &  6.16 & $ 1.2 \pm 0.2$ & $  +3 \pm 9  $ & ...           & $1.4^{+0.2}_{-0.3}$   & $ 8.7^{+0.3}_{-0.4}$ & $80^{+10}_{-37}$ \\

& & & & & & & & \\

\multicolumn{9}{c}{\textbf{Southern Region-- Dec.\,(J2000)\,$<-$9:00:38.30}} \\

6   & 80.73 & 64.62  & $0.7  \pm 0.1$ & $ +56 \pm 7  $ & $ -16 \pm  7$ & $1.5^{+0.1}_{-0.2}$   & $ 8.5^{+0.3}_{-0.1}$ & $81^{+9}_{-41}$  \\
12 & 81.26 &  8.45   & $0.4  \pm 0.1$ & $ -12 \pm 3  $ & $-131 \pm 14$ & $2.0^{+0.2}_{-0.2}$   & $ 8.2^{+0.2}_{-0.1}$ & $90^{+46}_{-46}$ \\
16 & 79.50 & 17.14  & $ 2.1 \pm 0.3$ & $ -22 \pm 8  $ & $ -31 \pm 12$ & $1.6^{+0.2}_{-0.3}$   & $ 9.0^{+0.2}_{-0.3}$ & $84^{+6}_{-22}$  \\
27 & 79.37 &  4.45   & $ 1.8 \pm 0.2$ & $ +89 \pm 3  $ & $+151 \pm 12$ & $<0.5$                & $ 8.9^{+0.1}_{-0.1}$ & $89^{+1}_{-15}$  \\
38 & 82.84 &  1.90   & $2.3  \pm 0.2$ & $ -44 \pm 3  $ & $ -77 \pm 13$ & $<0.5$                & $ 9.0^{+0.1}_{-0.2}$ & $90^{+12}_{-12}$ \\
55 & 70.28 &  0.61   & $ 6.2 \pm 0.8$ & $  -7 \pm 3  $ & $-155 \pm 21$ & $<0.5$                & $ 8.7^{+0.2}_{-0.5}$ & $90^{+8}_{-8}$   \\
58 & 82.49 &  1.77   & $3.5  \pm 0.7$ & $ -49 \pm 2  $ & ...           & $<0.5$                & $ 8.7^{+0.2}_{-0.5}$ & $90^{+12}_{-12}$ \\

 & & & & & & & & \\
\hline
 & & & & & & & & \\
\end{tabular}

\tablefoot{Labels in Col.\,1 are used to associate the current measurements with cloudlets identified in Table\,4 of
\citet{Sanna2010b}. Columns\,2 and~3 report the LSR velocity and brightness of the brightest spot of each cloudlet
on 2011 May 29. Columns\,4 and~5 give the measured linear polarization fraction and the position angle (east of north)
of the linear polarization vectors for each cloudlet, respectively. Column\,6 gives the position angle of the velocity
vectors as measured from \citet{Sanna2010b}, for comparison with column\,5. For each maser, Cols.\,7, 8, and~9
give the model results for the emerging brightness temperature, the intrinsic thermal linewidth, and the angle
between the magnetic field and the maser propagation direction, respectively.\\
\tablefoottext{a}{New feature, see Fig.\,\ref{fig2}.}
}

\end{table*}
}

\end{document}